\documentstyle[twoside,fleqn,espcrc2,epsf]{article}

\newcommand{\beq}{\begin{equation}}
\newcommand{\eeq}{\end{equation}}

\title{Reconstructing the conformal mode in simplicial gravity}

\author{S. Catterall\address{Physics Department, 
        Syracuse University, \\ 
        Syracuse, NY 13244}%
        and E. Mottola
        \address{Theoretical Division T-8, Mail Stop B285, \\
        Los Alamos National Laboratory, Los Alamos, NM 87545}}
       
\begin{document}

\begin{abstract}
We verify that summing 2D DT geometries correctly reproduces the Polyakov
action for the conformal mode, including all ghost contributions, at large
volumes. The Gaussian action is reproduced even for central charges
greater than one lending 
strong support to the hypothesis that the space of all possible
dyamical triangulations approximates well the space of physically
distinct metrics independent of the precise nature of the matter
coupling.
\end{abstract}

\maketitle

\section{Quick Review of 2DQG}

For 2DQG coupled to matter fields $\phi$ with
central charge $c_m$ the path integral is:
\begin{equation}
Z=\int\frac{DgD\phi}{\mathrm{Vol\left(Diffs\right)}}
e^{-S\left(g,\phi\right)}
\end{equation}
It is important to include only physically distinct metrics. This
can be accomplished by fixing the gauge
$g=\overline{g}e^{2\sigma}$. This yields
\begin{equation}
Z\sim\int D\sigma e^{-S_L\left(\sigma\right)}
\end{equation}
The Liouville action $S_L\left(\sigma\right)$ is given by
\begin{equation}
S_L\left(\sigma\right)=\frac{25-c_m}{24\pi}\int\sqrt{\overline{g}}\left(
-\sigma\Box\sigma+\overline{R}\sigma\right)
\end{equation}
Equivalently the same result can be derived from the
trace anomaly of massless fields in a curved background \cite{emil}
\begin{eqnarray*}
T&=&\frac{\left(25-c_m\right)}{24\pi}R\\
&=&\frac{\left(25-c_m\right)}{24\pi}e^{-2\sigma}\left(\overline{R}-
2\stackrel{-}{\Box}\sigma\right)\\
&=&\frac{1}{\sqrt{\overline{g}}}\frac{\delta}{\delta
\sigma}S_L\left(\sigma\right)
\end{eqnarray*}
Thus the quantum effective action is the sum of this
anomaly-induced action and the classical action.

\section{
Consequences of $S_L\left(\sigma\right)$}

Notice that the quantum action contains non-trivial dynamics. Specifically
it ensures that 2D quantum gravity has the following properties:
\begin{itemize}
\item Nontrivial scaling of correlation functions
$\left<O_1\ldots O_N\right>\sim 
A^{p_1+\cdots +p_2}$.
\item Geometries are {\it fractal} eg $d_H=4$ for pure 2D gravity.
\item Baby Universe substructure  ($\gamma <0$).
\end{itemize}

\section{Branched Polymers}

For $c_m>1$ we find that, although the
action remains well-defined the 
scaling dimensions become
{\it complex}. At this point a
BKT-like argument due to Cates \cite{cates}
indicates that {\it spike} configurations dominate where
$\sigma_{\rm spike}\sim -\log{r}$. Such configurations have a
free energy which is sensitive to the U.V cut-off.
It is presently unknown whether the dominance of these
configurations indicates a complete breakdown of
Liouville theory or is simply a signal that it
is merely {\it incomplete} - perhaps the action should be
augmented by more terms whose couplings must be tuned to approach
a continuum limit. 

\section{
Dynamical Triangulations}

DTs furnish another approach to 2DQG - 
finite {\it simplicial } meshes are used to
approximate continuum geometries.
It is a fundamental postulate of this approach that
summing
over lattices generates 
the correct measure on the space 
of physically distinct metrics
\begin{equation}
Z=\sum_{T,\phi}e^{S\left(T,\phi\right)}
\end{equation}
where
\begin{equation}
S=\sum_{\left<ij\right>\epsilon T}\phi_i\phi_j+\cdots
\end{equation}

The strongest evidence for this comes from the startling
agreement
of correlation functions computed using Liouville theory
or via the DT approach \cite{amb}.

\section{
 Lattice conformal mode}

Given the observed agreement between the correlaton functions we
would
like to demonstrate explicitly the equivalence of the two
formalisms.
We shall assume
that every DT geometry can be thought of as approximating a continuum
metric which can be
conformally mapped to
the round sphere with constant curvature $\overline{R}$.
The conformal factor needed is given by
\begin{equation}
R=e^{-2\sigma}\left(\overline{R}-2\stackrel{-}{\Box}\sigma\right)
\end{equation}
Since $e^{-2\sigma}\stackrel{-}{\Box}=\Box$ this
can be rewritten as a non-linear lattice equation:
\begin{equation}
2M_{ij}\sigma_j=\frac{2\pi}{3}\left(6-q_i\right)-\frac{\overline{R}q_i}{3}
A_\Delta e^{-2\sigma_i}
\end{equation}
where
\begin{equation}
M_{ij}=\frac{2}{\sqrt{3}}\left(q_i\delta_{ij}-C_{ij}\right)
\end{equation}

\section{
Gaussian distribution}

For each DT triangulation generated in the
Monte Carlo simulation we solve this
equation. This
yields a distribution of the 
lattice conformal mode.
To check against Liouville we then
decompose the lattice field
on eigenmodes of
$\stackrel{-}{\Box}$.
On 
the lattice we have
\begin{equation}
L_{ij}=\frac{3}{A_\Delta}\frac{e^{\sigma_i+\sigma_j}}{\sqrt{q_iq_j}}M_{ij}
\end{equation}
where
\begin{equation}
L_{ij}u^\ell_j=\lambda^\ell u^\ell_i
\end{equation}
The
amplitude
of each mode
$\sigma^\ell=\sum_i\frac{q_i}{3}
e^{-2\sigma_i}u^\ell_i\sigma_i$ should then be distributed according to
\begin{equation}
\exp{-\frac{\left(25-c_m\right)}{24\pi}\sum_l\lambda^\ell\left(\sigma^\ell\right)^2}
\end{equation}

\section{
Results}

\begin{figure}[htb]
\vspace{9pt}
\epsfxsize=2.25 in
\epsfbox{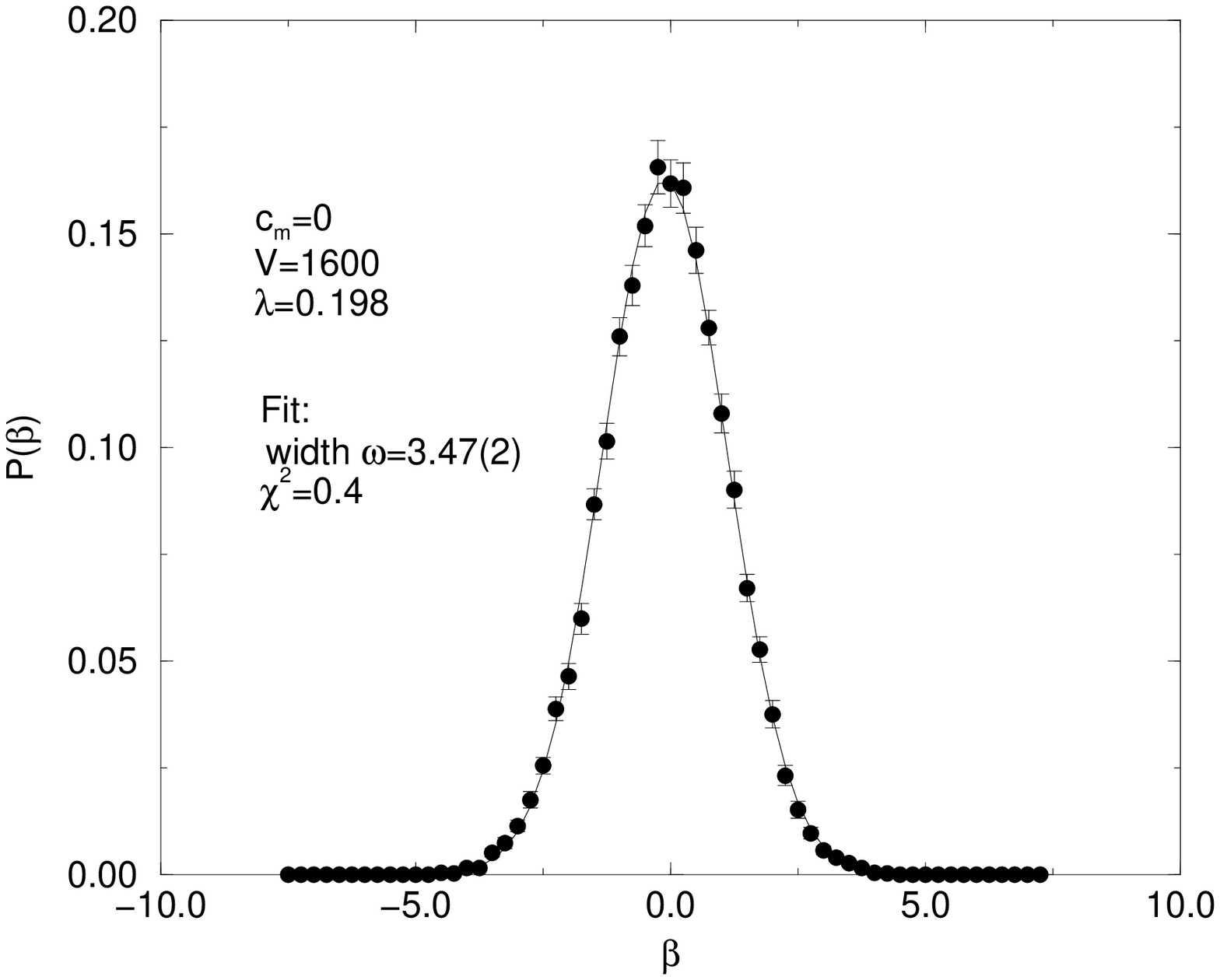}
\caption{Distribution of $l=10$ mode with gaussian fit for
$c_m=1$ and $V=1600$}
\label{fig1}
\end{figure}

\begin{figure}[htb]
\vspace{9pt}
\epsfxsize=2.25 in
\epsfbox{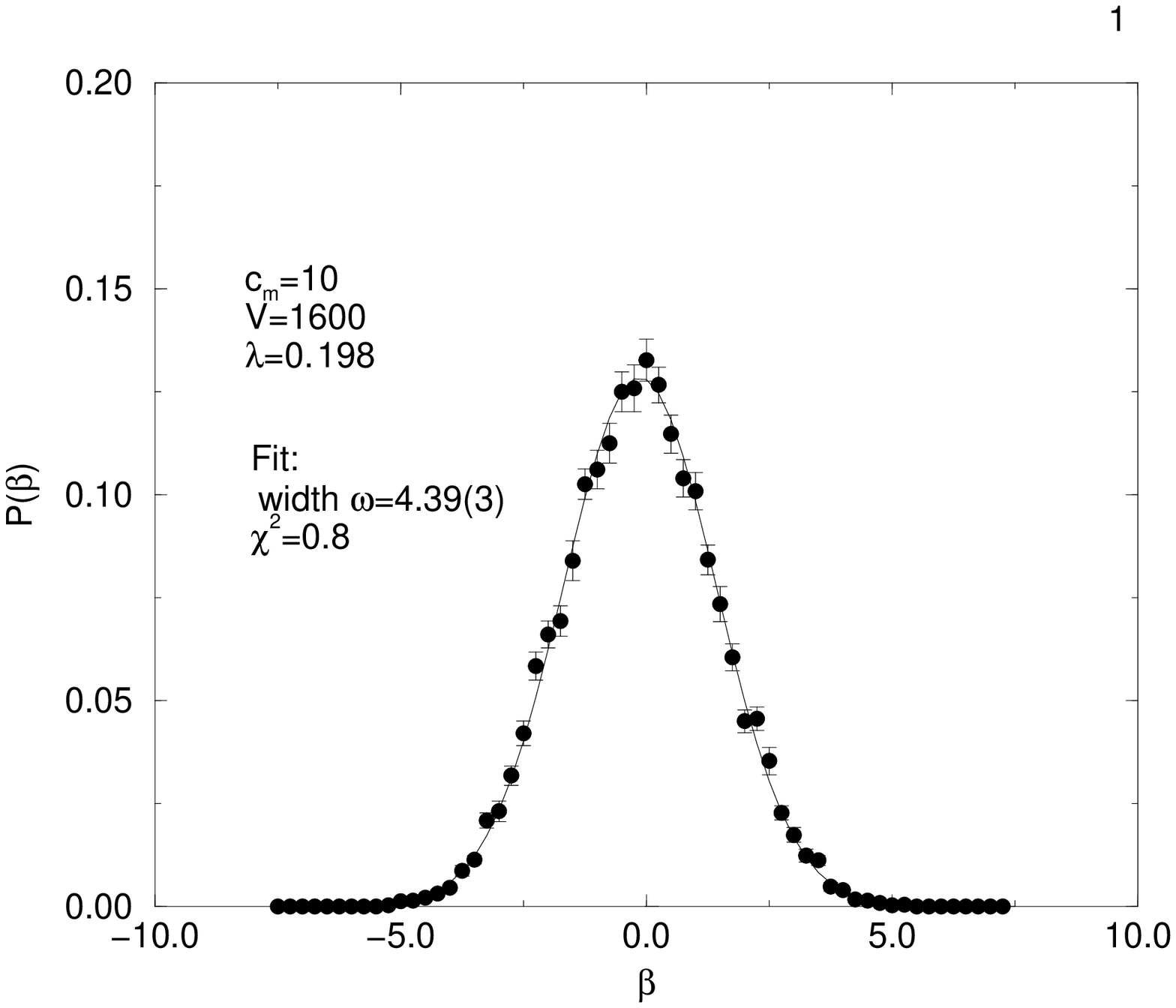}
\caption{Distribution of $l=8$ mode with gaussisan fit for
$c_m=10$ and $V=1600$}
\label{fig2}
\end{figure}

\begin{figure}[htb]
\vspace{9pt}
\epsfxsize=2.25 in
\epsfbox{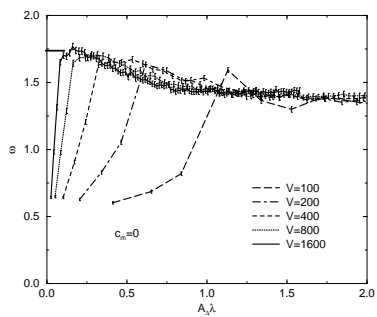}
\caption{Width versus lattice eigenvalue for $c_m=1$}
\label{fig3}
\end{figure}

\begin{figure}[htb]
\vspace{9pt}
\epsfxsize=2.25 in
\epsfbox{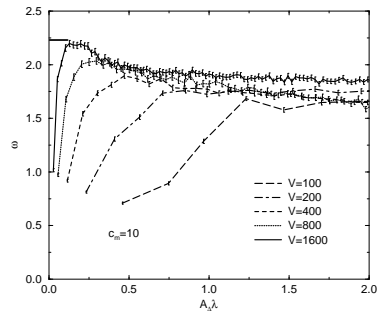}
\caption{Width versus lattice eigenvalue for $c_m=10$}
\label{fig4}
\end{figure}

First we
rescale
amplitudes
$\sigma^\ell\to\sigma^\ell
\sqrt{\lambda^\ell}$. Each mode $\ell$ should now be 
distributed with equal width depending only on the central charge.
Gaussian fits are then performed
to extract these widths
$\omega_\ell$. Results for $c_m=1$ and $c_m=10$ are shown in figures
\ref{fig1} and \ref{fig2}. The widths of the fits agree well
with Liouville theory. The mode dependence of the width is 
shown in fig.\ref{fig3} and fig.\ref{fig4} as a function of
lattice volume $V$. The lattice
eigenvalue $A_\Delta \lambda_\ell$ is plotted along the x-axis. 
In the
continuum limit $A_\Delta\to 0$ and $V\to\infty$
we see good
agreement with the Liouville
prediction for all
$c_m$ (for a more complete discussion see \cite{me})

Furthermore, the same arguments can be used to
analyze the 
zero
mode distribution
\begin{equation}
P\left(\sigma^0\right)
\sim\exp{-\frac{\left(25-c_m\right)}{12\pi V}\sigma^0}
\end{equation}
We have observed that the distribution of the zero mode follows this
theoretical prediction
for large argument (where the fixed area constraint plays no role)
and allows for an independent measurement of the central
charge. As for the nonzero modes this yields values that are statistically
consistent with the continuum value.

\section{
Conclusions}

It is possible to
recover the
conformal mode in 2D simplicial QG. It is
distributed according to
Polyakov-Liouville action with the correct central charge
including all ghost contributions. For $c_m<1$ this
amounts to anothing convincing test of the equivalance of DT
approach to continuum aproaches to 2D quantum gravity. Our results
for $c_m>1$ would further support the conclusion that the
DT measure is an appropriate measure for sampling the physical
space of geometries even when there are truly propagating matter
degrees of freedom.

\end{document}